\documentclass[twocolumn,twoside,slac_two]{revtex4}
\usepackage{graphicx}
\usepackage{fancyhdr}
\begin{document}

\title{Synchrotron Emission from Pair Cascades in AGN Environments}

%

\author{P. Roustazadeh and M. B\"ottcher}
\affiliation{Astrophysical Institute, Department of Physics and Astronomy, \\
Ohio University, Athens, OH 45701, USA}

\begin{abstract}
Recent detections of very-high-energy (VHE, $E > 100$~GeV) $\gamma$-ray blazars 
which do not belong to the high frequency peaked BL Lac (HBL) class, suggest 
that $\gamma\gamma$ absorption and pair cascading might occur in those objects. 
In the presence of even weak magnetic fields, these Compton-supported pair
cascades will be deflected and contribute to the {\it Fermi} $\gamma$-ray flux
of radio galaxies. We demonstrate that, in this scenario, the magnetic field 
can not be determined from a fit of the cascade emission to the $\gamma$-ray 
spectrum alone, and the degeneracy can only be lifted if the synchrotron 
emission from the cascades is observed as well. We illustrate this fact with the example of NGC 1275. 
We point out that the cascade synchrotron emission may produce spectral features reminiscent
of the big blue bump observed in the spectral energy distributions of
several blazars, and illustrate this idea for 3C~279.
\end{abstract}

\maketitle

\thispagestyle{fancy}

\section{Introduction}
Flat spectrum radio quasars (FSRQs) and BL Lacertae objects (BL Lac) are the two 
main subclasses of blazars. They radiate in the entire electromagnetic spectrum, 
from radio up to $\gamma$-rays. The spectral energy distribution (SED) of blazars 
consists of a low energy peak and a high energy peak. It is strongly belived that the low energy component 
from the radio to optical-UV or x-rays is dominated by synchrotron radiation from
relativistic electrons. In leptonic models, the high energy component from x-rays 
to $\gamma$-rays is produced by Compton upscattering. Soft photons for Compton 
upscattering may be produced by the accretion disk \citep{ds93,dsm92}, 
the broad line region (BLR) \citep{dss97,sikora94}, hot dust in the central 
region \citep{bla00} or synchrotron emission from the highly relativistic particles 
in the jets themselves.

If the VHE $\gamma$-ray emission is produced in the high-radiation-density
environment of the broad line region (BLR) and/or the dust torus of an
AGN (as commonly found in non-HBL blazars), it is expected to be strongly 
attenuated by $\gamma\gamma$ pair production. 
In \cite{rb10,rb11}, we considered the full 3-dimensional development of
Compton-supported VHE $\gamma$-ray induced cascades in the external radiation
fields in AGN environments. In this paper, we summarize the salient results of
our recent work in which we generalized the Monte-Carlo cascade code developed 
in \cite{rb11} to non-negligible magnetic fields and consider the angle dependent 
synchrotron emission from the cascades. 

\section{\label{setup}Model Setup and Code Description}

The general model setup used for this work is described in \cite{rb10,rb11}.
The primary VHE $\gamma$-ray emission is represented as a mono-directional
beam of $\gamma$-rays propagating along the X axis, described by a power-law 
with photon spectral index $\alpha$ and a high-energy cut-off at $E_{\gamma, max}$. 
We have verified that the shape of the cascade emission is virtually independent
of the precise shape of the primary $\gamma$-ray spectrum, so that our conclusions
remain unchanged when using a more realistic VHE $\gamma$-ray emission spectrum
(such as a broken power-law or a power-law + exponential cut-off).
We assume that the primary $\gamma$-rays interact via $\gamma\gamma$ absorption 
and pair production with an isotropic radiation field with arbitrary spectrum
within a fixed boundary, given by a radius $R_{\rm ext}$.

\begin{figure*}[t]
\centering
\includegraphics[width=135mm]{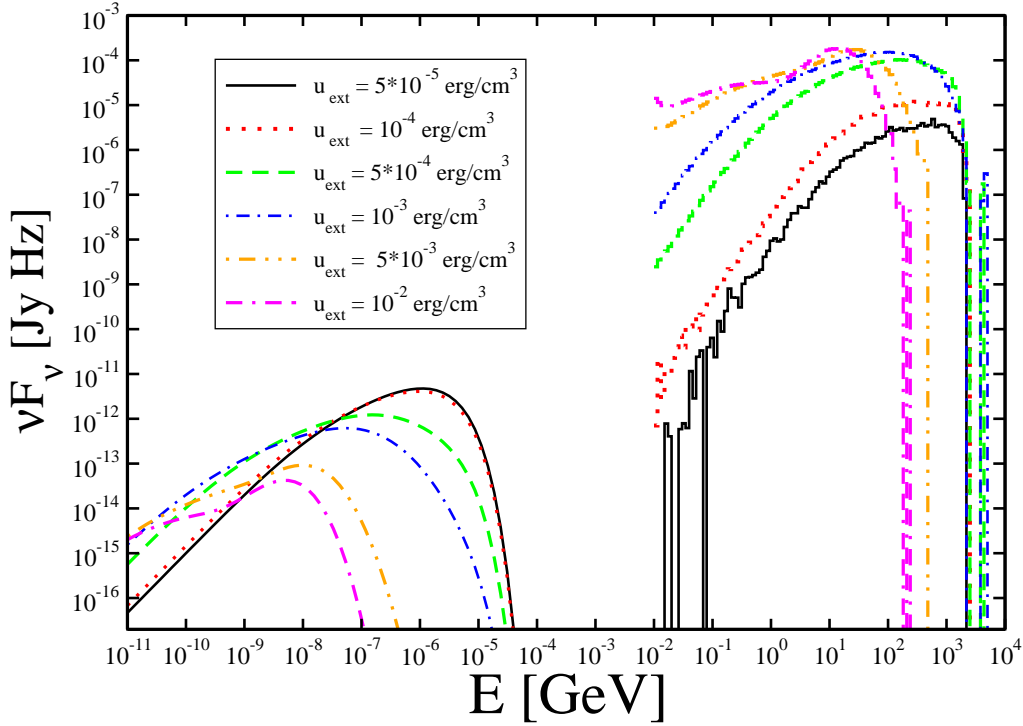}
\caption{\label{ufig}{The effect of a varying external radiation energy density.
Parameters: $B_x = 10^{-6}$~G, $B_y = 10^{-6}$~G, $\theta_B = 45^0$;
 $R_{\rm ext} = 10^{16}$~cm, $T = 1000$~K,
$\alpha = 2.5$, {$E_{\gamma, {\rm max}} = 5$~TeV} in the angular bin
$0.2\leq\mu\leq0.4$ } where $\mu$ is the cosine of the observing angle
with respect to the jet (x) axis.}
\end{figure*}

Our code evaluates $\gamma\gamma$ absorption and pair production using the
full analytical solution to the pair production spectrum of \cite{bs97} under
the assumption that the produced electron and positron travel initially along
the direction of propagation of the incoming $\gamma$-ray. The trajectories of 
the particles, as they are deflected by the magnetic field, are followed in 
full 3-D geometry. Compton scattering is evaluated using the head-on
approximation. The code calculates the synchrotron energy loss of cascade 
particles between successive Compton scatterings. The synchrotron emission
of electrons/positrons between two Compton scatterings is evaluated using
an asymptotic aproximation, $P_{\nu} (\gamma) \propto \nu^{1/3} exp(-\nu/\nu_c[\gamma])$,
properly normalized to the energy loss, where $ \nu_c \sim 3.4 \times 10^6 (B/G) \gamma^2 Hz$
and $\gamma$ is the electron Lorentz factor.

\section{\label{parameterstudy}Numerical Results}

We have used the cascade Monte-Carlo code described in the previous section
to evaluate the angle-dependent Compton and synchrotron spectra from VHE
$\gamma$-ray induced pair cascades.

\begin{figure*}[t]
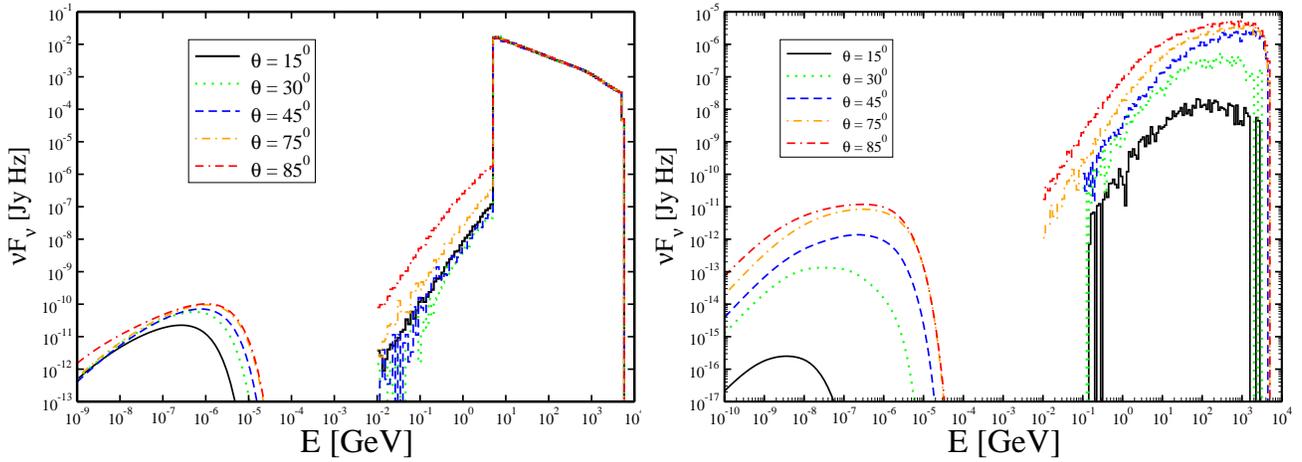

\vspace{1cm}
   \centerline{
        \includegraphics[width=85mm]{f5a.eps}
        \includegraphics[width=85mm]{f5b.eps}}
        \caption{\label{BfigF}{The effect of a varying magnetic field orientation
for a fixed magnetic field strength of $B = 1 \, \mu$G, $u_{\rm ext} =
10^{-6}$~erg~cm$^{-3}$, $R_{\rm ext} = 10^{18}$~cm, $T = 1000$~K, $\alpha = 2.5$,
{$E_{\gamma, {\rm max}} = 5$~TeV}. Left figure: (for angular bin $0.8\leq\mu\leq1.0$;
forward direction, blazars); Right figure: (for angular bin $0.2 \le \mu \le 0.4$,
radio galaxies)}}
    \end{figure*}
    
Figure \ref{ufig} shows the cascade spectra for different values of the external 
radiation field energy density $u_{\rm ext}$. For increasing values of the external 
radiation field, the spectral breaks of both radiation components, when viewed at 
substantial off-axis angles, shift to lower energies. This is a result of the 
decreasing Compton cooling length, implying that particles cool to lower energy
before being sufficiently deflected to contribute to the emission at a given
observing angle. Figure \ref{ufig} also shows that the synchrotron
luminosities of the cascades decrease with increasing $u_{\rm ext}$ while
the Compton luminosities of the cascades increase. For larger values of
$u_{\rm ext}$ and fixed blackbody temperature the soft target photon number
density increases and so does the photon-photon absorption opacity $\tau_{\gamma\gamma}$, 
so that an increasing fraction
of VHE photons will be absorbed, and the Compton flux from the cascades becomes 
larger. For very large values of $u_{\rm ext}$, $\tau_{\gamma\gamma}\gg 1$ for 
photons above the pair production threshold so that essentially all VHE photons 
will be absorbed and the Compton flux from the cascade becomes independent of 
$u_{\rm ext}$ \citep{rb10,rb11}. At the same time, the relative power in synchrotron
to Compton emission continues to be determined by $P_{\rm sy} / P_{\rm C}
\propto B^2 / u_{\rm ext}$. 

Figure \ref{BfigF} illustrates the effects of a varying magnetic-field orientation
with respect to the jet axis, for fixed magnetic-field strength $B = 1 \, \mu$G
for different angular bins. The results for the Compton component have been
discussed in \cite{rb11}, where have shown that the perpendicular component 
of the magnetic field ($B_y$) is responsible for the isotropization of secondaries 
in the cascade. The figure illustrates that also the synchrotron radiation depends
primarily on $B_y$.

\section{\label{degeneracy}Magnetic Field degeneracy}

\begin{figure*}[t]
\centering
\includegraphics[width=135mm]{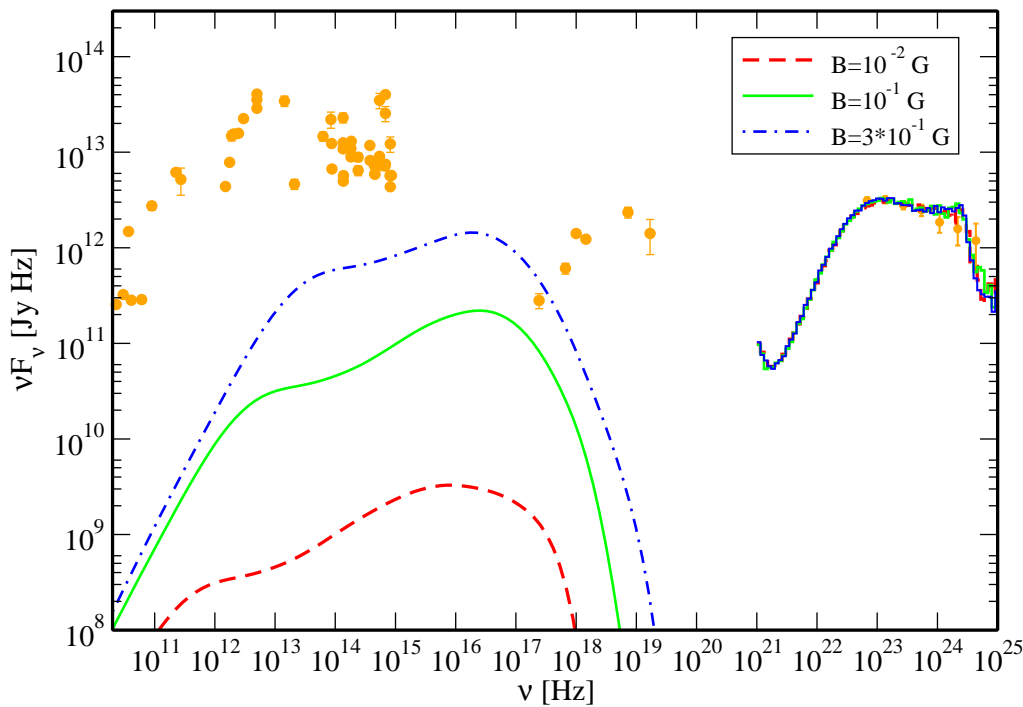}
\caption{\label{Degeneracy1}{
Synchrotron and Compton emission form the cascades for NGC~1275
($0.6 \leq\mu\leq 0.8 $). Parameters: $\theta_B = 11^o$;
$u_{\rm ext} = 5\times10^{-2}$~erg~cm$^{-3}$, $R_{\rm ext} = 10^{16}$~cm,
$E_s=E_{L\alpha}$,
$\alpha = 2.5$, {$E_{\gamma, {\rm max}} = 5$~TeV}.}}
\end{figure*}

In \cite{rb10}, we evaluated the exptected Compton emission from VHE $\gamma$-ray
induced pair cascades in the radio galaxy galaxy NGC~1275 (which is known to host
a low-luminosity AGN at its center), assuming that it would appear as a $\gamma$-ray 
blazar when viewed along the jet. This was used to fit the \emph{Fermi} spectrum 
of this radio galaxy. We now show that the parameter choice of the magnetic field, 
both orientation and strength, is degenerate in this fitting procedure, if only 
the high energy output from the cascades is considered. Figure \ref{Degeneracy1} 
illustrates this effect. In this plot, the external radiation field is
parameterized through $u_{\rm ext} = 5 \times 10^{-2}$~erg~cm$^{-3}$ with photon
energy $E_s = E_{Ly\alpha}$ and $R_{\rm ext} = 10^{16}$~cm. This size scale is
appropriate for low-luminosity AGN as observed in NGC~1275 \citep[e.g.][]{kaspi07},
and the parameters combine
to a BLR luminosity of $L_{\rm BLR} = 4 \pi R_{\rm ext}^2 \, c \, u_{\rm ext} =
1.9\times 10^{42}$~erg~s$^{-1}$, in agreement with the observed value for NGC~1275.
The magnetic field orientation is at an angle of
$\theta_B = 11^o$ with respect to the jet axis. The cascade spectrum shown in Figure \ref{Degeneracy1} pertains
to the angular bin $0.6 < \mu < 0.8$ (corresponding to $37^o \lesssim
\theta \lesssim 53^o$), appropriate for the known orientation of NGC~1275. In
\cite[]{rb10,rb11}, we have shown that for magnetic field values of $B_y\geq 1$~nG
and for an energy density of $u_{ext} \geq 10^{-3}$~erg~cm$^{-3}$, there is no 
pronounced break in the cascade spectrum and the cascade is independent of the 
magnetic field. In general, we expect no break in the cascade Compton emission 
if $E_{\rm IC,br} \gtrsim \frac{(mc^2)^2}{E_s}$ (where $E_{\rm IC, br}$ would 
be the expected spectral break where the Compton cooling length of the radiating
electrons is equal to their Larmor radius), which leads to the condition:

\begin{equation}
B_y \gtrsim \frac{{(m_e c^2)}^2 4\sigma_T u_{\rm ext} \theta}{3 e (E_s)^2} \sim 5 \,
u_{ext,-3} E_{s,1}^{-2} \theta \; {\rm nG}
\label{relation}
\end{equation}

Figure \ref{Degeneracy1} shows that while the high energy emission due to
deflection of the cascade up to the $\gamma\gamma$ absorption trough remains
virtually unchanged for different magnetic fields, the synchrotron emission 
from the cascade changes. Therefore, determining the B field requires knowledge 
of the synchrotron emission.

\section{\label{3C279}The Big Blue Bump}

3C~279 was among the first blazars discovered as a $\gamma$-ray source with
the Compton Gamma-Ray Observatory \citep[]{hartman92}. In 2007 it was detected
as a VHE $\gamma$-ray source with the MAGIC I telescope, making it the most
distant known VHE $\gamma$-ray source at a redshift of $0.536$ \citep{HB93}.
Its relativistic jet is oriented at a small angle to the line of the sight
of $< 0.5^0$ \citep[]{J04}. It is also detected by \emph{Fermi} \citep{abdo09c}
with photon spectral index $2.23$. There is evidence of a spectral break of
around a few GeV to a photon spectral index of $2.50$. It is strongly believed
that the radio to optical emission is due to synchrotron radiation by relativistic
particles in the jet. However, the origin of the high energy emission is still not
well understood \citep[see, e.g.,][]{br09}.

\cite{pian99} monitored 3C~279 in the ultraviolet, using IUE, and combined
their data with higher-energy observations from ROSAT and EGRET from 1992 December
to 1993 January. During this period, the source was in a very low state, allowing
for the detection of a UV excess (the BBB), which is typically hidden below a
dominant power-law continuum attributed to non-thermal emission from the jet.
\cite{pian99} proposed that the $\gamma$-ray emission in the SED of 3C~279
is produced by the external Compton mechanism, and suggested that the observed
UV excess might be due to thermal emission from an accretion disk.

\begin{figure*}[t]
\centering
\includegraphics[width=135mm]{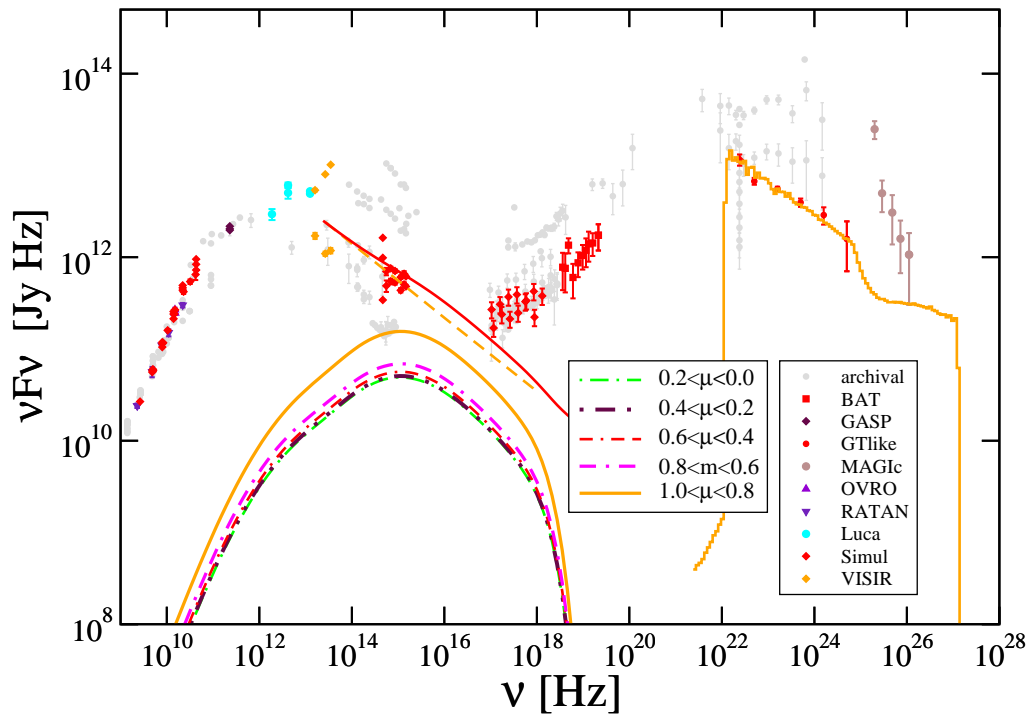}
\caption{\label{fit3C279}Illustration of a possible BBB in 3C~279 from
cascade synchrotron emission.
Parameters: $B= 10^{-2}$~G, $\theta_B = 85^0$;
 $R_{\rm ext} = 5\times10^{17}$~cm, $T = 2000$~K,
$\alpha = 2.37$, $u_{\rm ext} = 10^{-4}$~erg~cm$^{-3}$, {$E_{\gamma, {\rm max}}
= 5$~TeV}}
\end{figure*}

Here we propose an alternative model to explain the BBB in blazars. Figures 
\ref{ufig} -- \ref{Degeneracy1} illustrate that the synchrotron emission from 
VHE $\gamma$-ray induced pair cascades may peak in the UV/X-ray
range, thus mimicking a BBB for sufficiently strong magnetic fields
($B \gtrsim 1$~mG). Figure \ref{fit3C279} illustrates a possible BBB in 3C~279
produced by synchrotron emission from cascades, added to a phenomenological
power-law IR -- optical continuum attributed to jet synchrotron emission.

We suggest that synchrotron emission from VHE $\gamma$-ray induced pair
cascades can be an alternative explanation of the BBB in the SEDs of several
blazars such as 3C~279. An observational test of this hypothesis may be
provided through spectropolarimetry. A BBB due to (unpolarized) thermal
emission from an accretion disk will produce a decreasing percentage of
polarization with increasing frequency throughout the optical/UV range.
In contrast, if the BBB is produced as synchrotron emission from cascade
pairs in globally ordered magnetic fields, it is also expected to be polarized.
Therefore, we predict that a BBB due to cascade synchrotron emission would result
in a degree of polarization showing only a weak dependence on frequency over the
optical/UV range. 

\section{\label{summary}Summary}

We investigated the magnetic-field dependence and synchrotron emission
signatures of Compton-supported pair cascades initiated by the interaction
of nuclear VHE $\gamma$-rays with arbitrary external radiation fields in
the near-nuclear radiation environments of AGN. 

We demonstrated that when interpreting the $\sim$~GeV $\gamma$-ray flux
from radio galaxies (i.e., mis-aligned blazars) as cascade Compton emission
from VHE $\gamma$-ray induced pair cascades, the magnetic field can not be 
well constrained by considering the high-energy (Compton) output from the 
cascade emission alone, without observational signatures from their synchrotron 
emission.

We have shown that synchrotron emission from VHE $\gamma$-ray induced pair
cascades may produce UV/X-ray signatures resembling the BBB observed in the
SEDs of several blazars, in particular in their low states, and demonstrated
this with the example of 3C~279. We point out that spectropolarimetry may serve
as a possible observational test to distinguish a thermal from a non-thermal
(cascade) origin of the BBB.

\begin{acknowledgments}
This work was supported by NASA through Fermi Guest Investigator
Grants NNX09AT81G and NNX10AO49G.
\end{acknowledgments}

\bigskip 

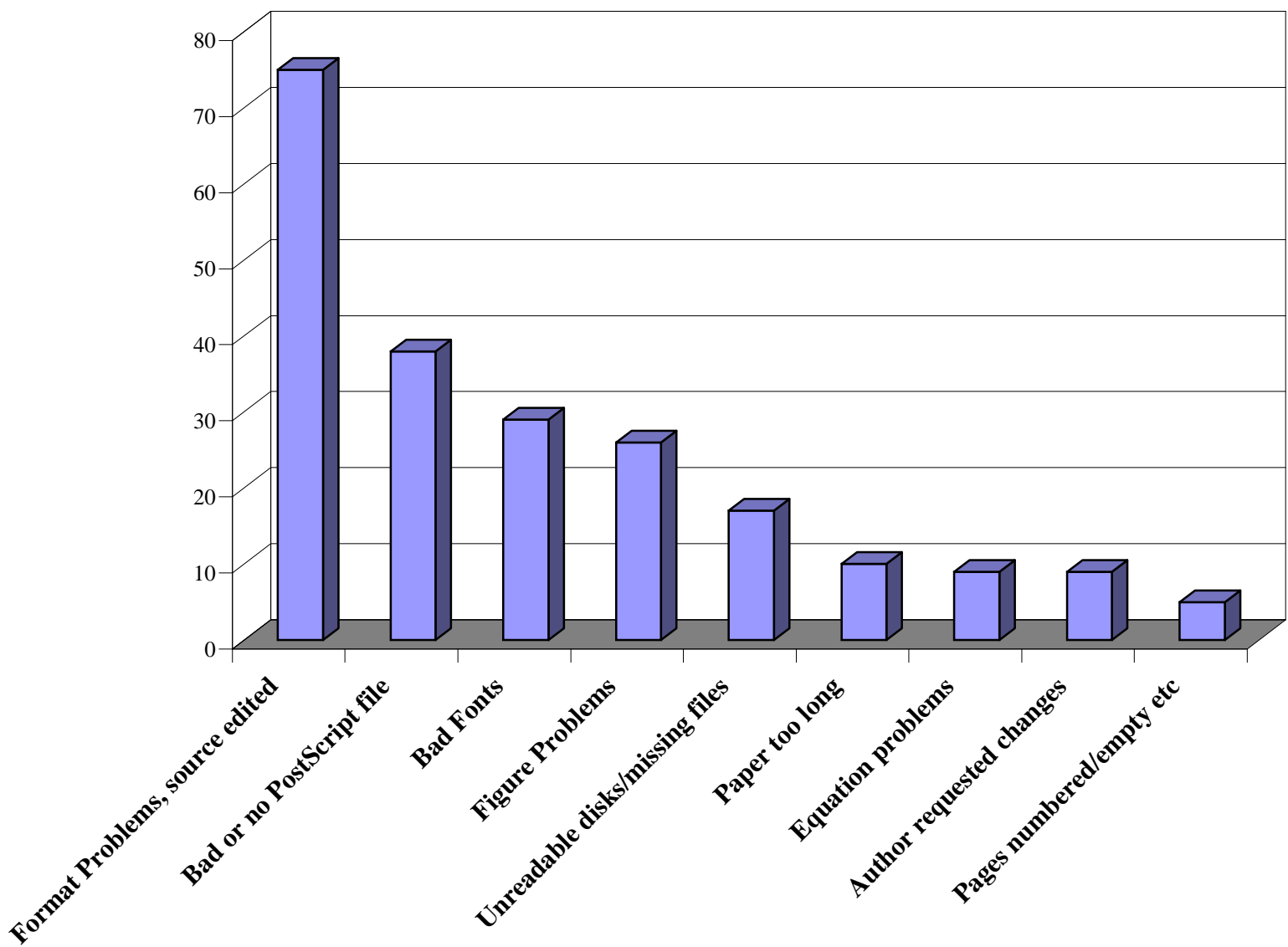
\end{document}